\documentclass{PoS}
\usepackage{amsmath,amssymb,graphics}
\usepackage{grffile}
\usepackage{mcite}

\graphicspath{{}}
\usepackage{array}
\newcolumntype{L}[1]{>{\raggedright\let\newline\\\arraybackslash\hspace{0pt}}m{
#1}}
\newcolumntype{C}[1]{>{\centering\let\newline\\\arraybackslash\hspace{0pt}}m{#1}
}
\newcolumntype{R}[1]{>{\raggedleft\let\newline\\\arraybackslash\hspace{0pt}}m{#1
}}
\usepackage{graphics,epsfig}
\usepackage{epstopdf}
\fontencoding{T1}
  \fontfamily{arial}
  \fontseries{m}
  \fontshape{n}
  \fontsize{20}{18}
  \selectfont

\newcommand{\tr}[0]{\text{tr}}

\usepackage{grffile}
\title{Complete Monopole Dominance of the Static Quark Potential}

\ShortTitle{Monopole Dominance}
\author{\speaker{Nigel Cundy}\\
        E-mail: \email{ndcundy@gmx.com}}

\abstract{
In earlier work, we used a gauge independent Abelian Decomposition to show that  Abelian degrees of freedom are wholly responsible for the static quark potential. The restricted Abelian field can be split into two terms, a Maxwell term and a $\theta$ (Dirac) term. The $\theta$ term's contribution to the string tension can be analysed theoretically and numerically, and arises because of the existence of a certain type of monopole. While the Abelian field can be constructed without gauge fixing, its two component parts are gauge-dependent, with a gauge transformation moving the topological features from one part to another. This allows us to isolate and identify the topological objects responsible for confinement by constructing a gauge where the $\theta$ term wholly accounts for the string tension. We confirm the presence of these monopoles in lattice simulations of SU(2) Yang-Mills theory. 
}
\FullConference{34th annual International Symposium on Lattice Field Theory\\
		24-30 July 2016\\
		University of Southampton, UK}

\begin{document}
\section{Introduction}
We wish to identify the topological objects that lead to quark confinement. We use the gauge-independent Abelian decomposition first proposed by Cho, Duan and Ge~\cite{Cho:1980,*Cho:1981,*F-N:98,*Shabanov:1999,*Duan:1979,*Kondo:2008su,*Shibata:2009af,*Kondo:2010pt,*Kondo:2005eq,*Shibata:2007pi}, which decomposes the non-Abelian SU($N$) gauge field $A_\mu$ into an Abelian (restricted) field, $\hat{A}$ and a coloured field $\hat{X}$,
\begin{align}
A_\mu(x) =& \hat{A}_\mu(x) + {X}_\mu(x) & U_{\mu,x} = & \hat{X}_{\mu,x}\hat{U}_{\mu,x} ,
\end{align}
where $U_\mu$ is the lattice gauge link $P[e^{ig\int A_\mu dx_\mu}]$, $P$ indicates path ordering, and $\hat{U}$ and $\hat{X}$ the links corresponding to the $\hat{A}$ and $X$ fields.
The decomposition is performed regardless of the choice of gauge by choosing an SU($N$) field $\theta$ at each location, and a subsequent colour direction $n_j = \theta \lambda_j \theta^\dagger$, where the index $j$ runs over the diagonal Gell Mann matrices $\lambda_3, \lambda_8, \ldots$. 
\begin{gather}
\hat{A}_{\mu} =\frac{1}{2} n_j \tr \left(\lambda_j \theta^\dagger A_\mu \theta + \frac{i}{g} \lambda_j \theta^\dagger \partial_\mu \theta\right).
\end{gather}
We call $\tr \lambda_j \theta^\dagger A_\mu \theta$ the Maxwell part of the restricted field and $\tr i \lambda_j \theta^\dagger\partial_\mu\theta$ the $\theta$ part.
The field strength can be expressed solely in terms of the colour direction $n_j$, which means that there are $N-1$ redundant degrees of freedom in $\theta$: $n$ is invariant under transformations $\theta \rightarrow \theta e^{i d^j \lambda_j}$, which suggests that we adopt the following parametrisation of $\theta$
\begin{gather}
\theta = \left(\begin{array}{cccc}
                  \cos a_1&i\sin a_1 e^{ic_1}&0&\hdots\\
                  i \sin a_1 e^{-ic_1}&\cos a_1&0&\hdots\\
                  0&0&1&\hdots\\
                  \vdots&\vdots&\vdots&\ddots
                 \end{array}\right)  
                 \left(\begin{array}{cccc}
                  \cos a_2&0&i\sin a_2 e^{ic_2}&\hdots\\
                  0&1&0&\hdots\\
                  i \sin a_2 e^{-ic_2}&0&\cos a_2&\hdots\\
                  \vdots&\vdots&\vdots&\ddots
                 \end{array}\right)\ldots
                  e^{i\left( d_j \lambda_j\right)}\label{eq:deftheta},
\end{gather}
with $0 \le a_i \le \frac{\pi}{2}$ and $c_i, d_i \in \mathbb{R}$.
If we choose $\theta$ so that under a gauge transformation $\Lambda$ (where  $U$ transforms as $U_{\mu,x} \rightarrow \Lambda_x U_{\mu,x} \Lambda^\dagger_{x+\hat{\mu}}$), $\theta_x \rightarrow \Lambda_x \theta_x $, then $\hat{U}$ and $\hat{X}$ will transform gauge covariantly,
\begin{align}
\hat{U}_{\mu,x} \rightarrow & \Lambda_x \hat{U}_{\mu,x} \Lambda^\dagger_{x+\hat{\mu}} & \hat{X}_{\mu,x} \rightarrow \Lambda_x \hat{X}_{\mu,x} \Lambda^\dagger_{x}.\label{eq:gaugeFix}
\end{align}
The $U(1)^{N-1}$ field $\hat{A}$ transforms as expected for an Abelian Field. In SU(2),
\begin{gather}
\frac{1}{2} \tr \left(\lambda_3 \theta^\dagger A_\mu \theta + \frac{i}{g} \lambda_3 \theta^\dagger \partial_\mu \theta\right)\rightarrow \frac{1}{2} \tr \left(\lambda_j \theta^\dagger A_\mu \theta + \frac{i}{g} \lambda_j \theta^\dagger \partial_\mu \theta\right) + \frac{1}{g}\partial_\mu d_3
\end{gather}
This means that we can construct gauge invariant observables from the restricted field alone. Since $\hat{U}$ is Abelian, it is much easier to treat it analytically. In particular, I want to study the static quark potential and the Wilson Loop. In~\cite{Cundy:2015caa,*Cundy:2013xsa} we showed that we can choose $\theta$ so that the restricted field wholly accounts for the Wilson Loop: by selecting $\theta$ as the matrix of eigenvalues of the path ordered product of gauge links starting and ending at that point. This means that the cause of confinement contributes only to the restricted field $\hat{A}$.
 The Wilson Loop resembles,
\begin{gather}
W_L = \tr \prod U_{\mu,x} = \tr e^{i g \lambda_j \oint dx_\mu  \frac{1}{2} \tr \left(\lambda_j \theta^\dagger A_\mu \theta + \frac{i}{g} \lambda_j \theta^\dagger \partial_\mu \theta\right)}.\label{eq:1.6}
\end{gather}
In terms of the parameters that describe the $\theta$ field, we have in SU(2),
\begin{gather}
\theta^\dagger\partial_\mu \theta =  i \partial_\mu a_1 \phi  + i \sin a_1 \cos a_1 \bar{\phi}\partial_\mu c_1 - i \sin^2 a_1 \partial_\mu c_1 \lambda^3 \label{eq:dtheta},\\
\phi =  \left(\begin{array}{cc} 0&e^{ic_1}\\e^{-ic_1}&0\end{array}\right) \phantom{hellobaby} \bar{\phi} =  \left(\begin{array}{cc} 0&ie^{ic_1}\\-ie^{-ic_1}&0\end{array}\right), 
\end{gather}
In particular, the expression for the Wilson Loop contains the factor
\begin{gather}
\tr W_L = \tr \prod U_{\mu,x} = \tr  e^{\ldots + i  \oint dx_\mu \left(\partial_\mu (\sin^2 (a_1) c_1) - c_1 \partial_\mu (\sin^2 a_1)\right)}.\label{eq:confinement}
\end{gather}
The parameter $c_1$ is ill-defined at $a_1 = 0$ or $\pi/2$. If we adopt some polar coordinates centred around the singularity and the Wilson Loop is defined in the $xt$ plane, then we can choose $(t,x,y,z) = r(\cos \varphi_1,\sin\varphi_1\cos\varphi_2,\sin\varphi_1\sin\varphi_2\cos\varphi_3,\sin\varphi_1\sin\varphi_2\sin\varphi_3)$, and we can write $c_1=\nu_c \varphi_1$ where $0\le \varphi < 2\pi$ for winding number $\nu_c$. The value of the Wilson Loop will contain a contribution proportional to the sum of the winding number within the loop, and thus can be expected to scale with the number of these monopoles within the loop, proportional to the area of the loop.

My goal is to show that the topological objects discussed above are indeed present in the configuration and contribute to quark confinement. The problem is that $\theta$ is gauge dependent; so before we can identify any objects we need to fix the gauge. This presents a problem: whatever causes confinement should not be gauge dependent. The two terms contributing to $\hat{A}$ vary under a gauge transformation as follows:
\begin{align}
\tr( n^j  (g   A_\mu)) \rightarrow &\tr( n^j  (g   A_\mu   +i \partial_\mu (\Lambda^\dagger) \Lambda))\nonumber\\
\tr ({i} \lambda^j (\theta^\dagger \partial_\mu \theta))\rightarrow& \tr ({i} \lambda^j (\theta^\dagger \partial_\mu \theta - \theta^\dagger \partial_\mu (\Lambda^\dagger)  \Lambda \theta)) + i \partial_\mu d_3.
\end{align}
$\Lambda$ can be parametrised in the same way as $\theta$; it can also contain the same type of topological objects. So what a gauge transformation does is transfer the topological objects from the Maxwell part to the $\theta$ part in the restricted field. The objects themselves are not gauge dependent: the only thing that is gauge dependent is which part of the restricted field they contribute to. It is easy to search for the objects within the topological part of the action: we know what $\theta$ is, we can compute $a_1$ and $c_1$, and search for winding. 

So what we propose doing is gauge fixing so that the $\theta$ term alone can account for the string tension. It might be thought that the gauge fixing removes the advantages of using the gauge independent decomposition; however this is not the case. The Abelian restricted field is wholly responsible for confinement regardless of which gauge we are in. The monopoles are present in the restricted field regardless of which gauge we are in. All we are doing by the gauge fixing is moving them from the Maxwell part to the topological part: a tool to help us to isolate them from any background noise and observe them directly. 

A second objection to our line of research is that our Abelian decomposition and Gauge Fixing procedure both depend on the Wilson Loop. If one constructs the restricted field, fixed the gauge and so on, and then studies a different Wilson Loop, there would no longer be perfect agreement between the string tensions of the original and restricted gauge fields. Surely the cause of quark confinement should be general; confining all possible quarks and not just this one pair of quarks we happen to be studying? Thus, the objection runs, we might have found a nice trick, but it bears little relation to the real physics behind quark confinement. This objection, however, is naive. When studying quark confinement, we are not interested in the confining potential in the absence of quarks, but, rather, the gluonic features that confine an actual pair of quarks. The presence of the quarks will certainly affect the underlying gluonic vacuum; one ought not expect the topological features contributing to confinement to be unaffected by their presence. Thus after we have placed two quarks on our lattice, we cannot ask what confines two hypothetical and non-existent quarks in a different location, but how the gluonic field reacts to the two quarks we have in front of us to produce a confining potential. The features in the gluonic field which confine quarks should be dependent on the location of the quark, i.e. Wilson Loop dependent. Also, in~\cite{Cundy:2016ckz} we show that on the lattice it is possible to reparametrise the gauge field so that the parameters $\cos (2a_1)$, $c_1$ and $d_3$ describing $\theta$ are among the dynamical gauge field variables. Essentially, the Abelian decomposition is just a change in the basis used to represent the gauge fields. In the Absence of matter, we are free to choose whatever basis we please; but in the presence of matter we should select whatever representation of the gauge field which makes the study of the quark-gluon interaction easiest.  Our Abelian decomposition is just a tool to represent the gauge field in a way that makes obvious the cause of confinement of a particular set of quarks.

These proceedings are arranged as follows: in section \ref{sec:2}, I describe the gauge fixing procedure used to isolate the topological objects from the rest of the gauge field. In section \ref{sec:3} I give numerical results, and I conclude in section \ref{sec:4}. This work is based on~\cite{Cundy:2016ckz}
\section{Fixing the Gauge}\label{sec:2}
Our numerical results are taken from an SU(2) pure Yang-Mills lattice ensemble, generated at $\beta = 2.0$ with a Symanzik improved gauge action~\cite{TILW,*TILW3}. The lattice volume is $16^3 \times 32$, and we used 10 steps of stout smearing~\cite{Morningstar:2003gk,*Moran:2008ra} at parameters $\rho = 0.1$, $\epsilon = 0$ to smooth the field before computing any observables. Our ensemble contained 94 configurations. 

Because the gauge fixing is computationally expensive, we have only so far performed this initial calculation in SU(2). We used a set of nested Wilson Loops in the $xt$ plane to construct $\theta$ and to measure the string tension; these loops are then stacked on top of each other in the $y$ and $z$ directions. The result of this is that $\theta$ is loop-dependent: which topological objects contribute to confinement depends on the quarks we are attempting to confine. There are three necessary requirements for our gauge fixing: 1) $\theta$ should be differentiable in the continuum limit (which is a requirement of the underlying theoretical construction); 2) the $\theta$ term in the restricted field should account for the whole Wilson Loop; 3) The gauge fixing procedure should be free of ghost terms (otherwise one would have to weight the gauge-dependent observables by a Fadeev-Poppov determinant). The Abelian decomposition ensures that $\theta^\dagger_x \hat{U}_{\mu,x}\theta_{x+\hat{\mu}}$ is diagonal and gauge invariant, however its value depends on the particular choice of $d_3$ when parametrising $
\theta$ according to equation (\ref{eq:deftheta}). The first step is to choose $d_3$ so that $\theta^{\dagger}_x \hat{U}_{\mu,x}\theta_{x+\hat{\mu}}$ is as smooth as possible (i.e. its trace is maximised) for all gauge links which contribute to the Wilson Loop, and also the remaining gauge links in the $xt$ plane, and (in SU(2)) one other direction. A suitable discretisation of the  $\theta$ part of the restricted field is $\hat{U}^\theta_\mu = e^{i \lambda_j \tr (\lambda_j \theta^\dagger_x\theta_{x +\hat{\mu}})}$. We then fix the gauge  so that 
\begin{align}
\Delta_{\mu,x} = \tr(\hat{U}^\theta_{\mu,x}- \theta^\dagger_x \hat{U}_{\mu,x}\theta_{x+\hat{\mu}})(\hat{U}^\theta_{\mu,x}- \theta^\dagger_x \hat{U}_{\mu,x}\theta_{x+\hat{\mu}})^\dagger\label{eq:2.1}
\end{align}
is minimised. In SU(2), the gauge transformation matrix has three parameters per lattice site, and (\ref{eq:2.1}) leads to one condition per direction per lattice site, so in principle if we apply this rule in three dimensions the system is fully constrained and it might be possible to find a solution with $\Delta = 0$. In SU(3), identifying the topological part with $\theta^\dagger_x \hat{U}_{\mu,x}\theta_{x+\hat{\mu}}$ gives two conditions on the gauge transformation matrix per direction, so one needs four dimensions fix all eight parameters. 

\begin{figure}
 \begin{center}
  \begin{tabular}{cc}
  \includegraphics[width=6.5cm,height=4cm]{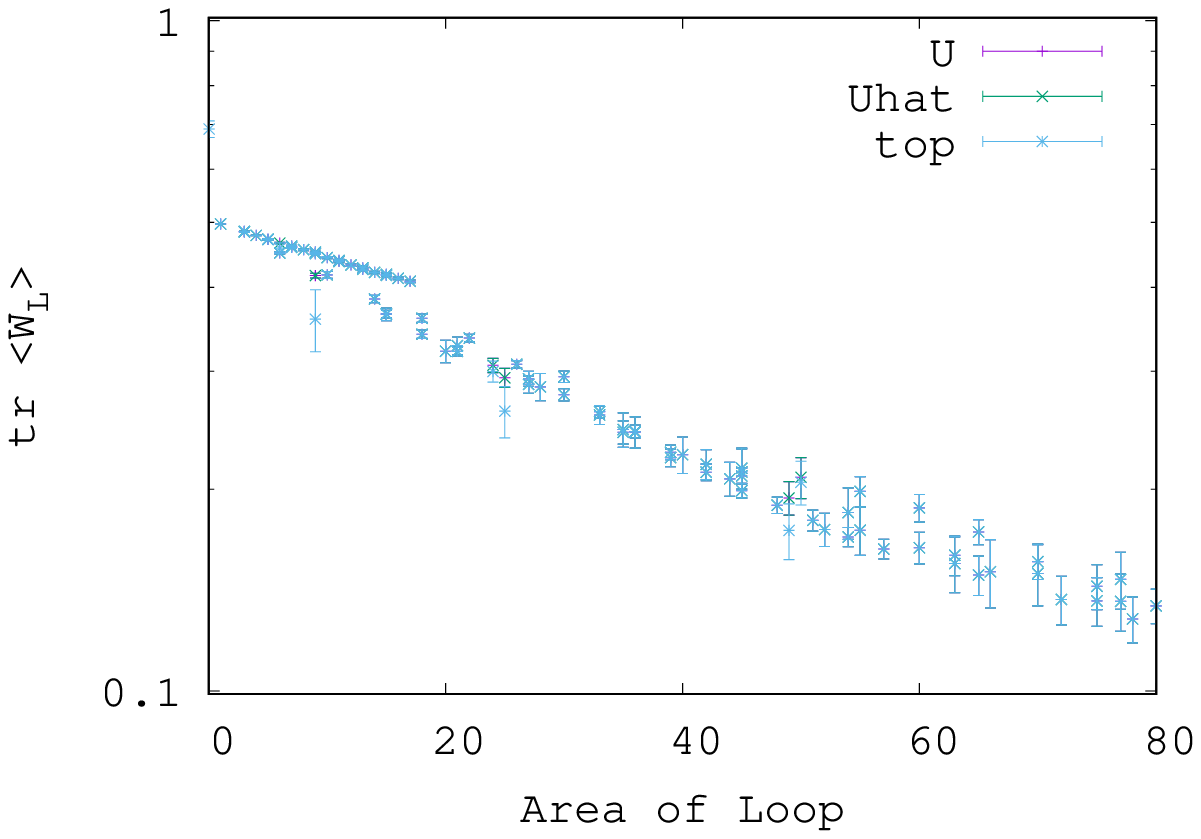}&
  \includegraphics[width=6.5cm,height=4cm]{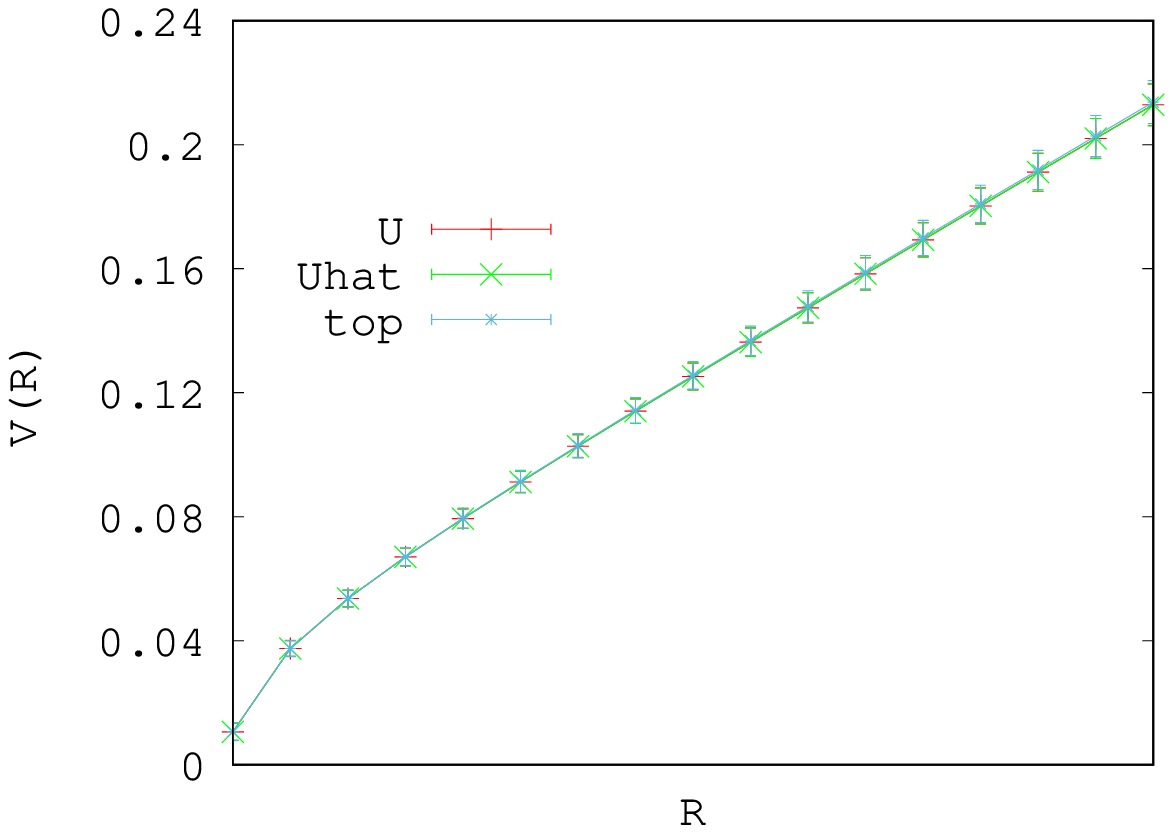}
  \end{tabular}
  \end{center}
 \vspace*{-.5cm}
\caption{The expected value of the Wilson Loop (left) and static quark potential (after an extrapolation to infinite time) (right), where the Wilson Loop is calculated with the Yang-Mills gauge field ($U$), the restricted gauge field ($Uhat$) and the topological part of the restricted gauge field ($top$).}\label{fig:WL}
\end{figure}

The condition with $\Delta_{\mu,x} = 0$ for all links along the Wilson Loop ensures that the $\theta$ part of the restricted gauge field completely dominates the string tension (see figure \ref{fig:WL}). The previous choice of fixing $d_3$ ensures that the $\theta$ field is smooth in the relevant directions. And, with some effort, one can also show that if $\Delta_\mu = 0$ in three dimensions, then this gauge fixing procedure has no Fadeev-Poppov ghosts (the Fadeev-Poppov determinant is constant). Unfortunately, our numerical experience was that we couldn't achieve $\Delta_\mu = 0$ in three dimensions; so we fixed it to zero for links contributing to the Wilson Loop and minimised it given that constraint in the two other directions.

This gauge fixing procedure achieved our goal of ensuring that the whole confining potential could be accounted for by the $\theta$ part of the restricted field, as seen in figure \ref{fig:WL}. With the full, restricted, and $\theta$ static potentials indistinguishable, we can begin the task of identifying the monopoles that lead to confinement.
\section{Identifying the Monopoles}\label{sec:3}
\begin{figure}
 \begin{center}
  \begin{tabular}{cc}
   \includegraphics[width=6.5cm,height=4.0cm]{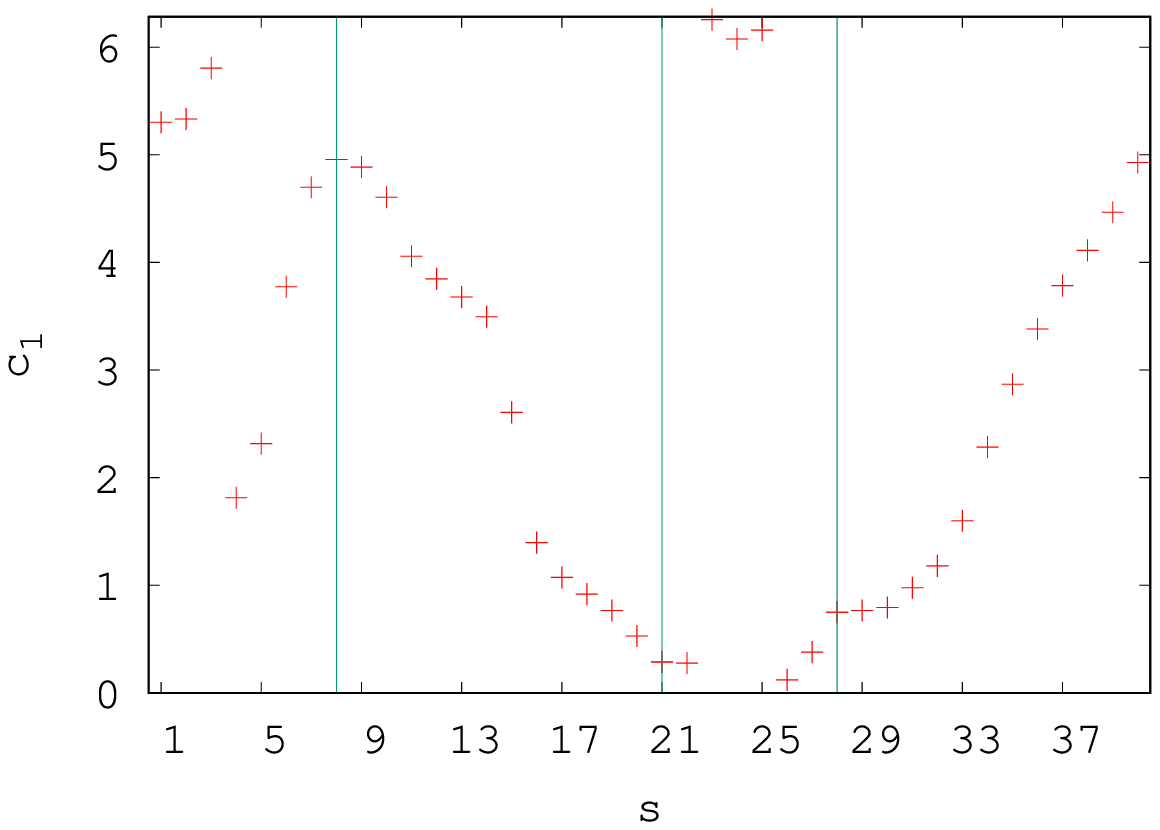}&
   \includegraphics[width=6.5cm,height=4.0cm]{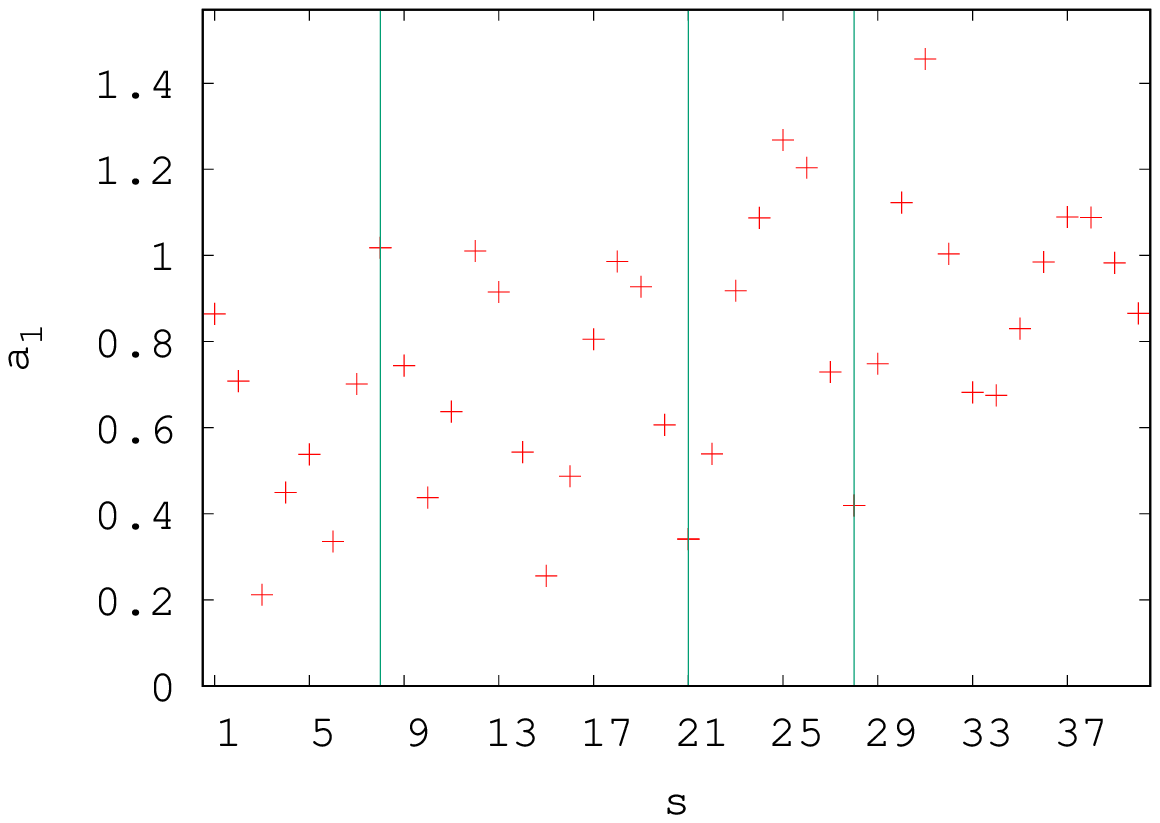}\\
   \includegraphics[width=6.5cm,height=4.0cm]{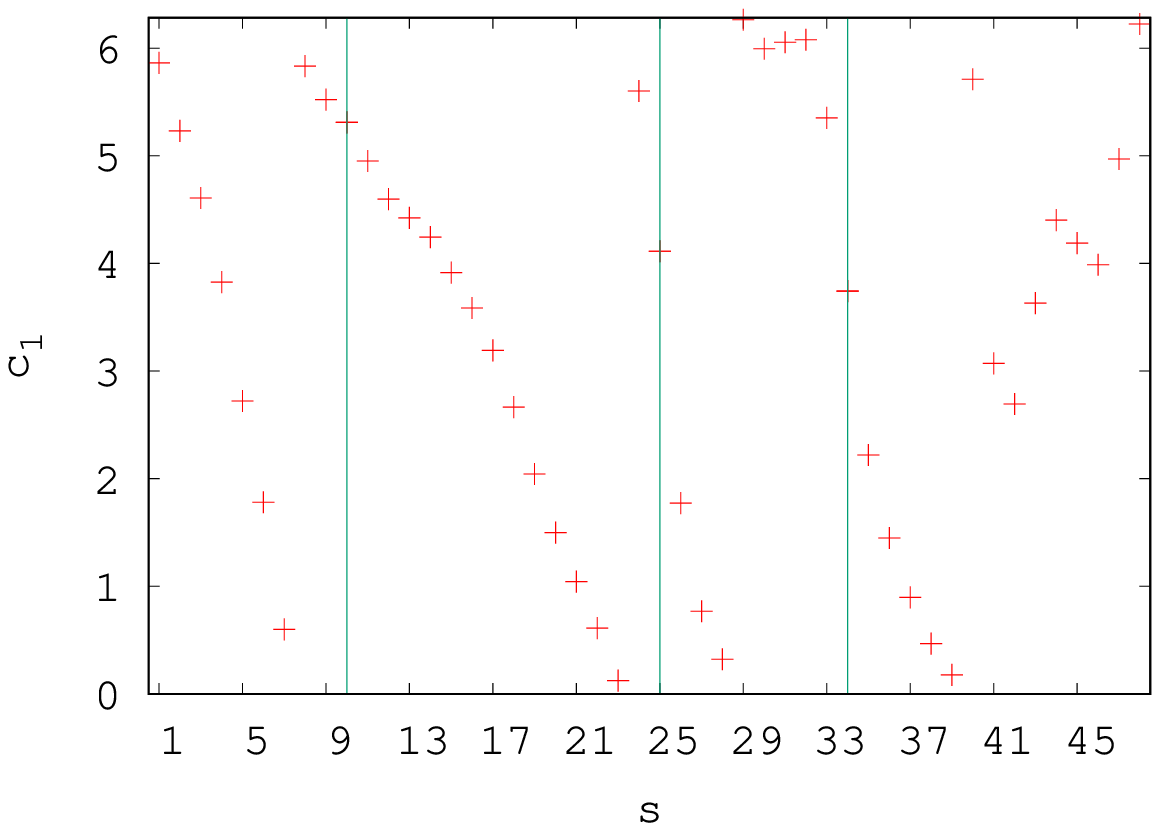}&
   \includegraphics[width=6.5cm,height=4.0cm]{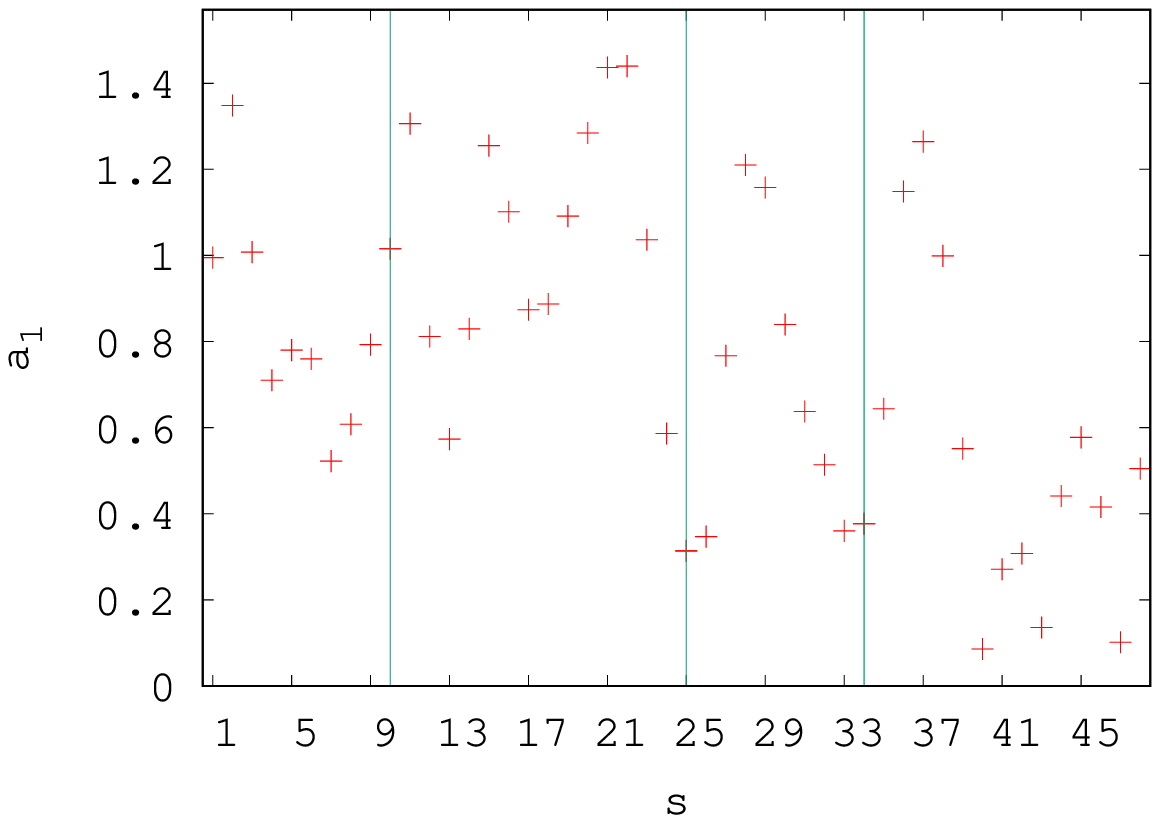}
  \end{tabular}
 \end{center}
 \vspace*{-.5cm}
  \caption{The parameters $c_1$ (left) and $a_1$ (right) extracted from the Abelian Decomposition $\theta$ matrix, after gauge fixing, shown along a $7\times 13$ rectangular Wilson Loop in the XT plane (top), and the $9\times 15$ loop surrounding it. The $x$-axis marks the position along the loop, starting at the lowest $x$ and lowest $t$ coordinates, and the vertical lines show the remaining three corners. Both axes of the plots are periodic, with the top of the plot equivalent to the bottom, and the left edge equivalent to the right. The top plot has a winding number $1$: $c_1$ gains $2\pi$ as it transverses the loop. The lower plot has winding number $-4$. The change in winding number was caused by minima in $a_1$ at positions $25,32,39,47$ in the lower plot and $3$ in the upper plot.}\label{fig:a_and_c}
 \end{figure}

Figure \ref{fig:a_and_c} plots the variation of the parameters $a_1$ and $c_1$ around a typical nested series of Wilson Loops. What these plots show is the emergence of non-zero winding numbers in $c_1$ as we increase the area of the loop, with the topological objects emerging as expected close to points where $a_1$ is at a minima. We can see large jumps in the parameter $c_1$ where $a_1$ is small or large; on the next loop this converts into a change in the winding number. This shows the emergence of the monopoles which we expect to contribute to confinement through equation (\ref{eq:confinement}). 

\begin{figure}
 \begin{center}
  \begin{tabular}{cc}
   \includegraphics[width=6.5cm,height=4cm]{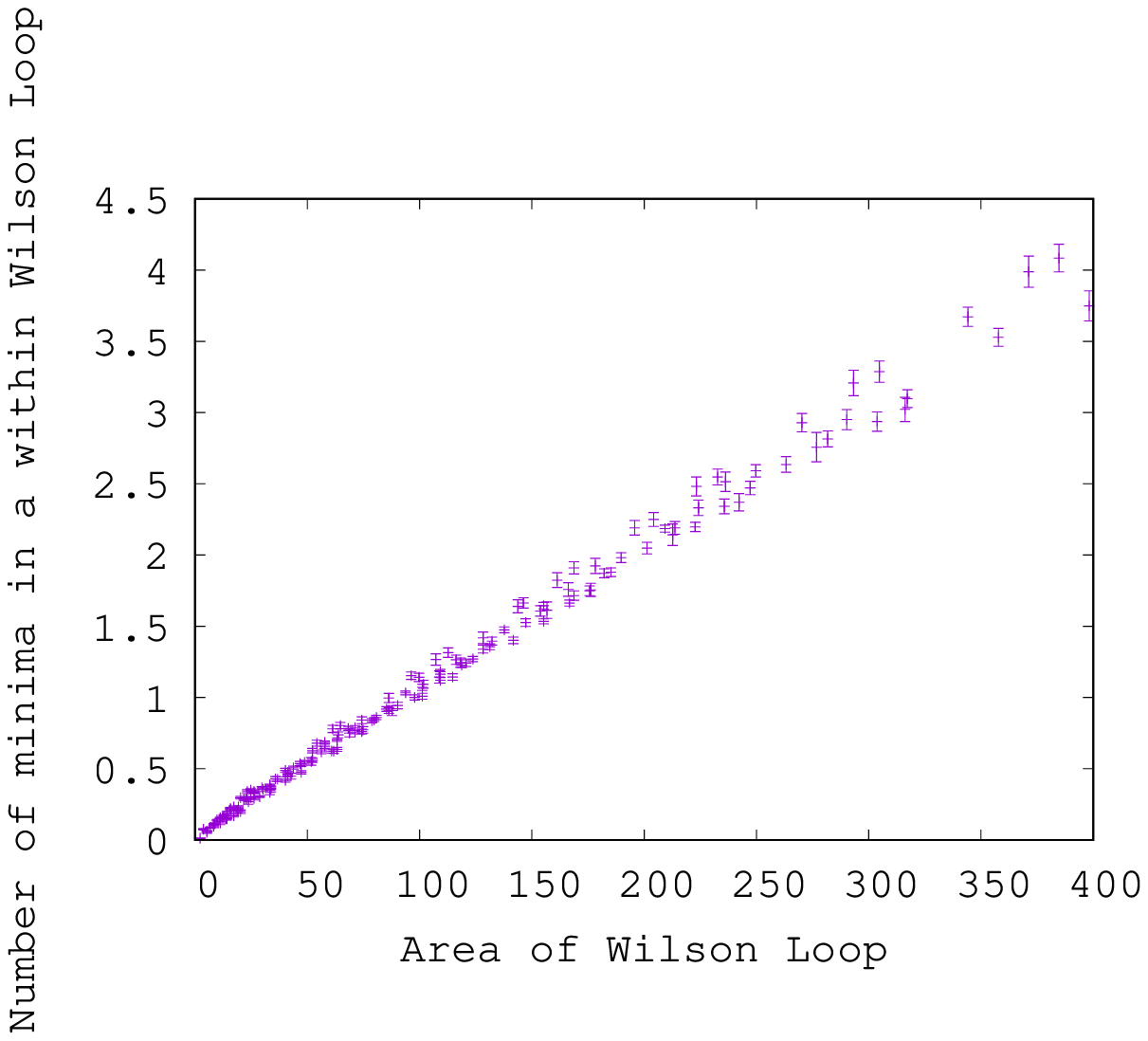} &
   \includegraphics[width=6.5cm,height=4cm]{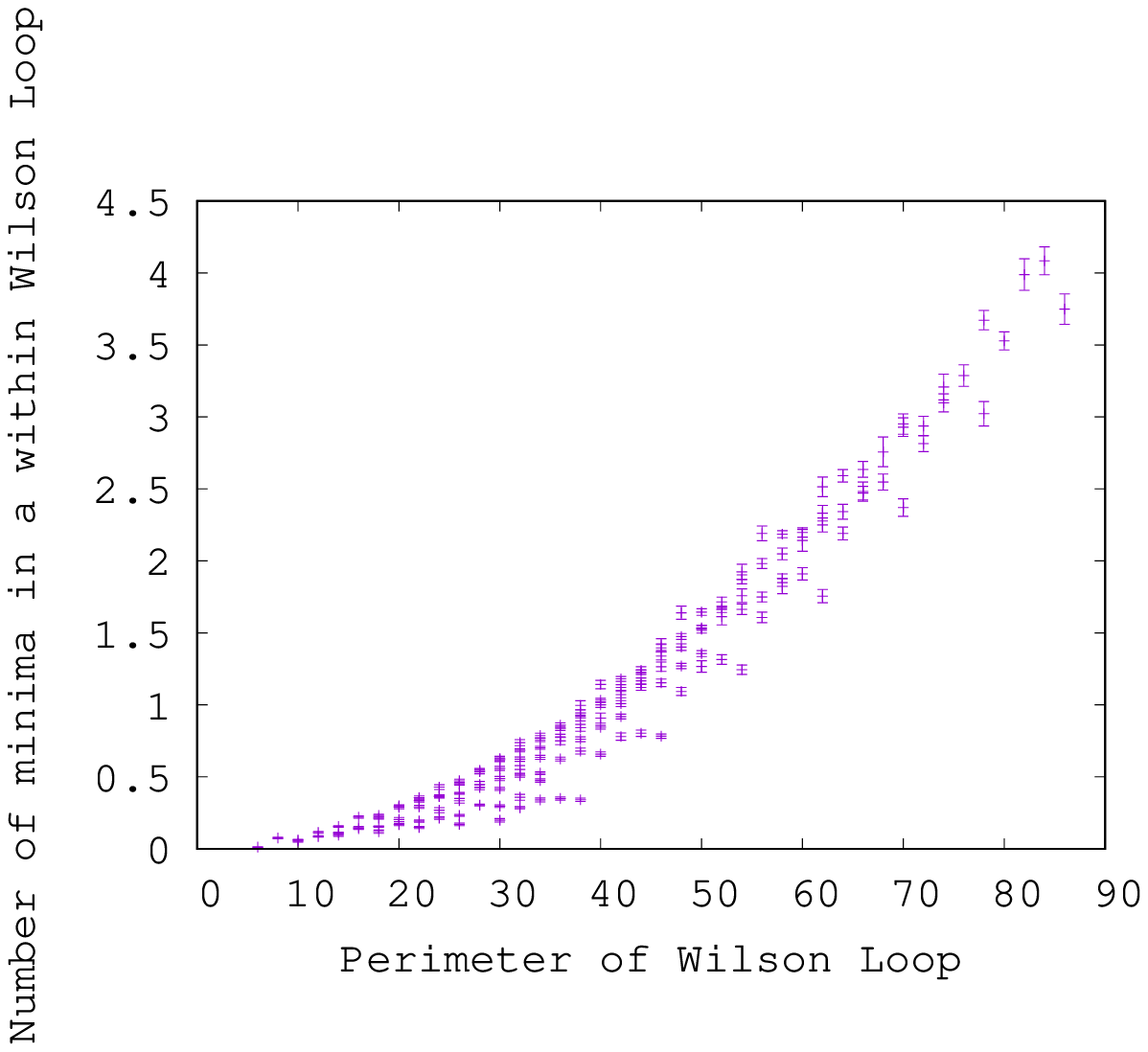} 
   \end{tabular}
 \end{center}
  \vspace*{-.5cm}
\caption{The average value of the number of minima of the parameter $a_1$ for those $\theta$ fields on lattice sites within the area bounded by the Wilson Loop plotted against the area of the Wilson Loop (left) and the perimeter of the Wilson Loop (right).}\label{fig:peaks}
\end{figure}

Do these objects obey an area or perimeter law? If they are to cause the observed string tension, the number of these objects would have to scale with the area of the Wilson Loop. The area law scaling is confirmed in figure $\ref{fig:peaks}$ where the number of monopoles is compared against both the area and perimeter of the Wilson Loop. 

\begin{figure}
 \begin{center}
  \begin{tabular}{cc}
   \includegraphics[width=6.5cm,height=4cm]{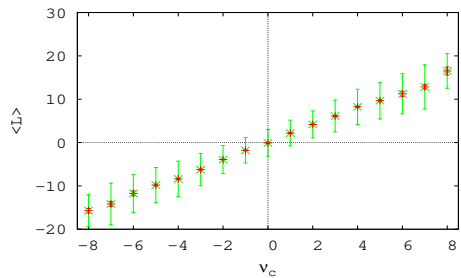}  
   \end{tabular}
 \end{center}
 \vspace*{-0.5cm}
\caption{The average value of the logarithm of the Wilson Loop plotted against the winding number. The smallest error bar is the error; the larger error bar is the standard deviation.}\label{fig:wind}
\end{figure}

Finally, we can ask whether the winding of the $c_1$ parameter is sufficient to account for the whole string tension. We define $L$ to be the exponent in equation \ref{eq:1.6},
\begin{gather*}
L = \frac{g}{2} \tr \left(\lambda_j \theta^\dagger A_\mu \theta + \frac{i}{g} \lambda_j \theta^\dagger \partial_\mu \theta\right).
\end{gather*}
If monopoles dominate the string tension, then this should be $\oint dx_\mu \sin^2 a_1 \partial_\mu c_1$, and with $\sin^2 a_1$ averaging to $\frac{1}{2}$ and the integral over $\partial_\mu c_1$ giving $2\pi$ times the winding number $\nu_c$, we expect on average $L$ to be proportional to $\pi \nu_c$. In practice, in a lattice simulation, discretisation effects will reduce the constant of proportionality; $\partial_\mu c$ is rendered in the lattice theory as $\frac{1}{\epsilon} \sin(c_{x+\epsilon \hat{\mu}} -c_x)$ where $\epsilon$ is the lattice spacing; and if the jump in $c_1$ from one lattice site to the next is large then the difference between these two expressions could be substantial. What we find (in figure \ref{fig:wind}) is that $L$ is proportional to the winding number, but the constant of proportionality is considerably smaller than $\pi$ (a linear fit gives gradient $1.99(2)$ with $\chi^2$ per degree of freedom of 1.6). We are investigating if this is a lattice artefact, or an issue with our gauge fixing procedure.
\section{Conclusion}\label{sec:4}
In this work we have shown that through a combination of Abelian decomposition and gauge fixing, we can isolate the topological objects responsible for confinement in SU(2) pure Yang Mills gauge theory. The $\theta$ term is wholly responsible for the static quark potential. We are able to identify the topological objects that correspond to a winding in the parameter $c_1$ that describes the $\theta$ field around the Wilson Loop. We have shown that these objects are indeed distributed according to an area law, and that the exponent of the Wilson Loop is proportional to the winding number.

 \section*{Acknowledgements}
 Computations were performed on servers at Seoul National University. 
 This research was supported by Basic Science Research Program through the National Research Foundation of Korea(NRF) funded by the Ministry of Education(2014063535).

  \bibliographystyle{JHEP_mcite.bst}
\bibliography{weyl}

\end{document}